# Microgrid Resilience:

# A holistic approach for assessing threats, identifying vulnerabilities, and designing corresponding mitigation strategies

Sakshi Mishra*, Kate Anderson, Brian Miller, Kyle Boyer, Adam Warren

National Renewable Energy Laboratory, Golden, CO
*corresponding author

Emails: Sakshi.Mishra@nrel.gov, Kate.Anderson@nrel.gov, Brian.Miller@nrel.gov, Kyle.Boyer@nrel.gov, Adam.Warren@nrel.gov

*Abstract*—Microgrids are being increasing deployed to improve the operational flexibility, resilience, coordinated-energy management capabilities, self-adequacy, and increased reliability of power systems. This strong market growth is also driven by advances in power electronics, improved control systems, and the rapidly falling price and increased adoption of distributed energy generation technologies, like solar photovoltaics and storage. In the event of grid outages, microgrids can provide a backup source of power; providing resilience to the critical loads; however, this requires that the microgrid itself is resilient to physical and cyber threats. Building highly resilient microgrids requires a methodological assessment of potential threats, identification of vulnerabilities, and design of mitigation strategies. This paper provides a comprehensive review of threats, vulnerabilities, and mitigation strategies and develops a definition for microgrid resilience. The paper also develops a methodology for designing resilient microgrids by considering how microgrid designers and site owners evaluate threats, vulnerabilities, and consequences and choose the microgrid features required to address these threats under different situations.

I. INTRODUCTION

    *A. Motivation and Background*

In a developed country like the United States, with an increasing shift toward an electrified and digital economy, many critical infrastructure sectors (e.g., communications, water, food, defense, healthcare) rely directly or indirectly upon the electric power supply. Thus, an uninterrupted and continuous supply of power is an essential component of continued well-being and sustainable development of the nation. For countries in the developing world, basic electrification plays a critical role in advancing economic development, which requires reliable power infrastructure. As decentralized generation sources are proliferating around the world, microgrids are becoming an effective means of generating and utilizing power locally.

The energy sector has been designated as a uniquely critical infrastructure "due to the enabling function [the energy sector] provide[s] across all infrastructure sectors" by Presidential Policy Directive 21 (PPD-21) [1]. Recent weather-related events and cyberattacks have brought the resilience of the energy sector to the forefront of the national research priorities. As stated by the Executive Office of the President, "the resilience of the U.S. electric grid is a key part of the nation's defense against severe weather" [2]. The economic and security impact of such events can be widespread beyond the stricken area.

Natural and human-induced disasters can result in loss of electricity, causing large financial losses that span multiple sectors and various service types, negatively impacting the day-to-day lives of thousands, if not millions, of residents during each incident. For example, water infrastructure (water treatment plants and pumps), as well as the transportation sector (traffic signals), are impacted when electricity supply is interrupted. Such interdependencies worsen the impact of an electricity outage on the economy as a whole. A 2012 Congressional Research Service study found that the inflation-adjusted cost of weather-related outages in 2012 is $25-billion to $70-billion annually [2].

The frequency and severity of power outages caused by extreme weather have been increasing [3] . A tropical storm can impose three phases of impacts on power infrastructure: 1) initial impact of wind and rain (Hurricane Hermine in 2016, which struck Florida.); 2) storm surge in coastal areas and near major inland waters (Hurricane Katrina in 2005, Lake Pontchartrain, Louisiana.); and 3) flooding due to precipitation (Hurricane Sandy in 2012, which struck the eastern United States) [4]. In the first phase, high-speed winds can bring down the overhead wires, whereas the storm surge in second phase can cause considerable damage to generation and storage assets of the microgrid which are located close to the coastline. The next phase of flooding after storm surge hampers the restoration process.

Other natural hazards such as earthquakes also have devasting impact on the grid. The potential for earthquake disruption of major power system equipment is significant. The 2018 Hokkaido Eastern earthquake knocked out power to 5 million people in Hokkaido, Japan. A 1.65-gigawatt power plant that supplies more than half of the power of the largest Japanese main islands was shut down following a magnitude 6.7 quake.

In 2018, at least 80 reports of weather-related power outages were reported in the United States, which affected at least 50,000 customers [5]. In 2013, the Executive Office of the President published a report [2] pointing to severe weather as the single most common cause of power outages in the United States, representing over half of all outages and 87% of all outages that affected 50,000 or more customers from 2002-2012. [2]. Hurricane Sandy caused power outages to which 50 deaths are attributed [2]. In 2008, over 129 line faults in Southern China occurred due to a snowstorm [6].

Beyond weather events, human-induced outages are also a concern. A gunman in Metcalf California hit 17 transformers, resulting in $15 million in equipment damage [6] [7]. A 2015 cyberattack on the Ukrainian grid caused a power outage for six hours and left approximately 225,000 people without power. The BlackEnergy malware attack infected a utility's network through spear-phishing email, then harvested the credentials required to gain access to their Supervisory Control and Data Acquisition (SCADA) system operation. The attacker then opened breakers to bring more than 30 substations offline [8]. More recently, a cyber event within the power grid was reported in the United States in 2019 [9]. These cyber events in the Ukraine and in small U.S. utilities may be an indication of adversarial preparations, increasing their readiness to conduct larger-scale attacks, if needed, at a strategic time.

Given a host of natural and human-induced threats and the complexity of power systems, building resilience in energy infrastructure is a challenging task. Power system resilience has been defined as "a grid which has four fundamental properties of resilience, namely anticipation, absorption, recovery, and adaptability after the damaging events" [10]. A recent review paper on power grid resilience [10] looked at how several organizations define resilience. While every organization has its own definition of resilience, there are some commonalities. In general, resilience is characterized by resistance to damaging events, the ability to absorb the impact of events without loss of service to customers, the recovery capabilities after an event, and the adaptive ability of the system to learn from an event and prepare more effectively for the next one. The damaging events are typically characterized as low-probability, high-impact that affect large regions over long durations.

Planning for and building resilience solutions in power systems require a comprehensive evaluation framework. The steps include: 1) understanding the types and severity of the event for a given geographical area; 2) defining resilience metrics relevant to the situation; 3) identifying a suitable methodology for system assessment during and after the event; and 4) calculating the fault consequences, including estimating the true cost of the lost load. Building resilience is categorized into two major ways: system hardening and operational strategies for resilience [6].

Based on the type of natural disaster to which a given geographic area is prone, various system hardening approaches can be employed, including upgrading poles with robust materials, elevating substations, managing vegetation, and undergrounding distribution lines. Building redundancy (meshed networks instead of radial ones) is another system-hardening practice that has been widely adopted for decades by utilities as a reliability measure for critical loads, such as hospitals. However, redundancy is ineffective against certain events (e.g., a fire in an underground vault can destroy both sets of redundant power lines contained therein). Moreover, redundancy for large proportions of a network may be uneconomical.

A major bottleneck with employing system-hardening strategies is capital investment. Valuing the benefits of resilience can help in understanding the cost-benefit implications of hardening the system and potentially strengthen the business case to do so. Operational strategies present an alternative to capital investments and present a wide range of opportunities for building grid resilience. From a utility perspective, approaches such as advanced control and protection schemes, automated fault location, isolation using Advanced Metering Infrastructure (AMI), and demand-side-management can be employed to make the grid more resilient.

Dense interconnection of transmission systems strengthens power networks because these networks enable the movement of economically generated power from generation to load while offering redundancy. However, during a severe event, these interconnections can lead to cascading blackouts or load-shedding. Intact portions of the grid are often connected radially to impacted or damaged parts, spreading the impact of the fault. In radial networks, the power cannot be rerouted from the point of generation to the intact portions of the grid if a connecting component such as a transmission line is damaged. Grid hardening strategies, such as undergrounding lines, will not help with such cascading failures. Although underground lines avoid trees and wind faults, they can be damaged by floods or earthquakes and are more difficult to repair than aerial lines.

Localized resilience solutions, such as microgrids, are less vulnerable to cascading power system blackouts. The U.S. Department of Energy (DOE) defines the microgrid as "a group of interconnected loads and distributed energy resources within clearly defined electrical boundaries that acts as a single controllable entity with respect to the grid. A microgrid can connect and disconnect from the grid to enable it to operate in both grid-connected and island-mode" [2]. Microgrids are increasingly being utilized as backup systems for reliability and resilience solutions. Microgrids have largely been adopted by military bases, hospitals, academic institutions, cities, and ports. According to a market study report by Wood Mackenzie [11], the cumulative capacity of microgrid installations is forecasted to reach 8.8 GW by 2024 due to favorable market factors.

Microgrids can play a key role in providing resilience at the community or neighborhood level. Microgrids are unique among other grid-hardening measures and technologies, in that they can be installed by the customer, rather than requiring deployment by the utility. Significant savings through risk reduction and expeditious recovery can be achieved by building resilience in community-level generation and storage infrastructure [12]. Along with boosting resilience, benefits of microgrid to the local community include potentially reduced cost, as the microgrid can reduce demand on the grid during peak hours and also participate in markets by doing energy arbitrage.

As stated in [2], "A key feature of a microgrid is its ability during a utility grid disturbance to separate and isolate itself from the utility seamlessly with little or no disruption to the loads within the microgrid." The ability to operate independently from the grid is particularly important during weather-related grid outages, when the loss of power lines or substations may prevent the restoration of power, even after the utility's generation capabilities are back online [2]. During post-event grid-connected operational mode, a microgrid's surplus electricity generation capability can be used for providing a new reliability service aimed at accelerating the restoration of supply to neighboring customers [13]. Strategic installation of networked microgrids in weak regions of the power system has been shown to help with the post-disaster recovery of the utility grid [3] [14].

While microgrids are increasingly being adopted as a resilience and reliability solution, their own vulnerabilities cannot be ignored. The physical vulnerabilities of distributed generation-based microgrids are relatively lower for weather-related events compared to the utility grid, as no transmission and fewer distribution lines are involved. The microgrid may be impacted by a weather event if it is in close proximity to the epicenter of the event, but there is lower probability of interruption of supply due to a natural disaster or a human-induced attack happening somewhere else in the utility grid. But improperly designed microgrids can be more vulnerable to communications failures and cybersecurity-related events. For example, power substations are designed with dual redundant communication networks, but many microgrids have not been so designed to mitigate single points of failure. Therefore, it is essential to identify the risks, understand the vulnerabilities, and design the microgrids with mitigation measures that assuredly make them resilient against physical, communications, and cybersecurity-related threats. Systematically addressing these vulnerabilities during the design phase of the microgrid along with disaster preparedness to operate the microgrid during the most needed times will make microgrids a truly reliable, resilient, and enduring solution.

### B. Literature Review

Many studies have been conducted in the literature of power systems (utility grid) resilience. A report from the Executive Office of the President describes the economic benefits of increasing the resilience of power systems to outages caused by weather events in the context of US critical infrastructure enhancement [2]. Jufira *et al.* present a review of power systems resilience definitions and quantitative assessment methodologies for measuring resilience [10]. A detailed description of the damage caused by weather related-events to power systems infrastructure and its impact on bulk power system energy supply capability is presented in [15]. Another report from the US Department of Energy emphasizes the benefits of integrating distributed energy resources to increase infrastructure resilience [16]. Key strategies for realizing power systems resilience are discussed in [17]. Ouyang et at. focus on a resilience assessment of power systems with respect to hurricanes [18]. A methodology for determining optimal power systems investment for building resilience is presented in [19]. A framework for load restoration in the utility grid is proposed in [6], and strategies for modeling large-scale energy infrastructure from a resilience perspective have been explored in the literature [20]. Power system's infrastructure resilience has therefore been studied extensively; but microgrid resilience remains relatively unexplored. There is a need for understanding and measuring the resilience of microgrids (in islanded mode or as standalone entities in developing nations) against various types of threats to make them effective backup systems.

The process of designing resilience microgrids starts with a risk assessment. A risk assessment is a systematic way of identifying possible threats and estimating the severity of the corresponding vulnerabilities. A variety of risk assessment methodologies are used in industry, including Preliminary Hazard Analysis (PHA), What-if, Hazard and Operability (Hazop) studies, Failure Model and Effect Analysis (FMEA), Fault Tree Analysis (FTA), Expected Damage-Cost Analysis (EDCA), and Quantitative Risk Analysis (QRA) [21]. For microgrid-deployment projects, several approaches have been taken. Williams et al. present a case study of assessing risks to a microgrid for rural electrification in developing nations [22]. A risk-based performance analysis of microgrids with distributed energy generation is presented in [23]. A risk assessment for incorporating mitigation measures to make a microgrid resilient with cybersecurity perspective is presented in [24]. Risk analysis for microgrid deployment, therefore, is a multifaceted process where first the business risks associated with economic operations are analyzed, and then the next step is to identify and quantify the risks associated with the resilience of the microgrid.

A quantitative framework for a microgrid resilience assessment against windstorms is presented in [25]. An operational strategy to cope with the adverse impacts of extreme windstorm are presented in [26]. In this work, when the windstorm forecast alert or a flooding alert is received, the microgrid's operational strategy is modified through network reconfiguration, demand-side resources, generation reschedule, optimal parameter settings of droop-controlled units (such as Combined Heat and Power), and conservative voltage regulation to help it ride through the event. Likewise, an operational strategy to sustain operations during flooding is presented in [27]. A similar proactive operational scheduling for building operational resilience against hurricanes for multi-carrier microgrids is presented in [28]. Liang et al. [29] propose a control strategy for electric springs[1] to enhance the operational resilience of microgrids. Recently, Liu et al. have researched the cybersecurity aspect of microgrid resilience and analyzed DC microgrid resilience under a denial-of-service threat [30].

These works have focused on one specific aspect of microgrid resilience at a time, including physical sturdiness from natural disasters and maintaining cybersecurity. The work presented in this paper encompasses a holistic qualitative approach for assessing the external threats and associated vulnerabilities to a microgrid, and provides design and operational strategies tailored for mitigating the risk associated with identified threats. This work details both physical (hardware) and controls (software) dimensions of the microgrid to design a strategy for building microgrids capable of addressing threats in physical, cyber, and communications dimensions. The specific contributions of this work are as follows:

- We identify and categorize various threats to power systems and microgrid operations, including physical, cyber, and communications threats. We also present a quantitative threat-modeling methodology.
- We determine microgrid vulnerabilities associated with the threats identified in the previous step.
- We propose various mitigation strategies for enhancing the microgrid design to be resilient in both the physical and controls dimensions. We also recommend operational strategies for minimizing damage during the disaster and for faster recovery of the microgrid afterward.
- As an extension of resilience, we also briefly cover the reliability aspects of microgrid performance and differentiate between the two.

### C. Article Structure

With Section I.A laying out the motivation behind this work and a detailed review of the microgrid resilience literature in I.B, the rest of the paper is organized as follows: Section II introduces the preliminaries of reliability and resilience in the context of a microgrid. The risk assessment methodology is defined in Section III. Section III.C.2) delineates the threats to the resilient performance of power systems, including microgrid operations, and identifies the vulnerabilities associated with the threats. An example case study for risk factor calculation is presented in Section V. Section VI presents various mitigation measures, which can be employed for protecting the microgrid against different kind of threats as a preventive measure and various strategies for decreasing post-disaster recovery time. The article is concluded in Section VII where future research directions are identified.

## II. RESILIENCE AND RELIABILITY

Power system reliability has been studied in detail in the literature. Standardized metrics are used across the industry to measure the reliability of electricity supply from the utility. However, the development of resilience metrics is an active research area, and industry-accepted metrics for measuring the resilience of microgrids (or the utility grid) do not yet exist. It is essential to describe the difference between resilience and reliability of a microgrid to make effective design choices addressing the two aspects appropriately.

### A. Reliability

Billinton et al. describe reliability in power systems as covering "all aspects of the ability of the power system to perform its intended function of providing an adequate supply of electrical energy to customers efficiently with a reasonable assurance of continuity and quality" [31]. DOE's definition of reliability focuses on the ability of the systems to withstand sudden disturbances, such as electric short circuits or unanticipated loss of system components [32]. Reliability concerns of the utility grid can be divided into two main categories: system adequacy and system security. The sufficiency of generation, transmission, and distribution resources is required for maintaining system adequacy—the assessment is associated with static system conditions. On the other hand, system security relates to the ability of the system to maintain the continuity of the supply by responding to perturbations arising within the system, which include transient and dynamic disturbances. System security also concerns the ability of components or equipment to perform efficiently for a particular period, under a specified condition.

Events undermining the reliability of power systems typically last seconds (transient or dynamics disturbances) or hours (an unplanned outage of a generation unit or a major transmission line) and are caused by N-1 or N-1-1 contingency conditions. There are standardized metrics, such as System Average Interruption Duration Index (SAIDI), System Average Interruption Frequency Index (SAIFI), and Customer Average Interruption Duration Index (CAIDI), developed by the North American Reliability Corporation (NERC) for assessing the reliability performance of the utility grid at the individual utility level. Traditional solutions used for ensuring the reliability of the utility grid include meshed grids on the distribution side (alternative feeders to restore the supply, components in parallel) and advanced protection devices. Distributed energy resources (DERs) and other smart grid technologies are providing new opportunities for maintaining system reliability in the grid context.

Similar aspects are associated with reliability for microgrids; however, the solutions to ensure reliability in the microgrid are not the same as the solutions typically employed for the utility grid. A microgrid has a constrained amount of generation resources, and there may not be redundancies in components for economic reasons. To increase its reliability, a heavy emphasis needs to be placed on the reliable operation of the components in the microgrid (i.e., testing and preventative maintenance). Considering the case of resilience, reliable operation of the microgrid components (without any external threat) is a *necessary but not sufficient* condition to endure an external threat. On the other head, if the reliability of the components is already weak, then adding mitigation measures to increase resilience of the microgrid is of little use since the foundation of the microgrid's operational performance (i.e. reliability) is itself weak.

---

[1] electric springs is a new smart grid technology introduced in [59].

As discussed in the previous paragraph, understanding the reliability of the components of the microgrid is also an aspect of the resilient performance. However, it is a small subset of the large resilience performance equation. Reliability can loosely be correlated with the probability of vulnerability of system components (vulnerability is described in detail in Section III.B) by asking the following question: how likely is it that components forming both physical and cyber layers of the microgrid will accomplish the functional objectives in the face of an external threat? Apart from the probability of component failure, the contingencies that may cause reliability concerns in the microgrid are: 1) communications failure on the utility grid during grid-connected operations; 2) internal communications failures causing a disconnect between the master microgrid controller and physical assets, for centralized microgrid topology; and 3) disconnect between the distributed controllers causing uncoordinated operation of various physical assets in a decentralized microgrid topology.

### B. Resilience

Resilience, on the other hand, is the ability of a system (and its components) to adapt to changing conditions; and withstand and recover from disruptive events. It is a wide-reaching concept that does not just impact one component at a time. Instead, resilience is the system's ability to endure and recover from low-probability, high-impact events, such as natural disasters and human-induced attacks, that may impact large geographic regions over longer durations.

Figure 1 shows the process diagram of resilience analysis and deployment. Building resilient microgrids is an iterative process. A microgrid can be designed with features that address the identified threats and vulnerabilities; however, new threats or vulnerabilities may be revealed over time, particularly when the microgrid is tested during an actual disaster. Depending on its performance during and after the disaster, the microgrid design can be upgraded to adapt to newly identified threats and vulnerabilities. This cyclic model is especially relevant for cyberthreats because the attack surfaces and mediums through which the cyber infrastructure of a microgrid can be damaged are growing with increasing interconnectivity of the electrical and smart communications systems. Therefore, when the previously unknown vulnerabilities in the cyber-analytics layers are exposed through either an external attack or an internal awareness, the microgrid should adapt to enhance resilience toward the newly identified cyberthreats and minimize or eliminate its associated vulnerabilities. The five attributes shown in green are explained in detail in Table 11.

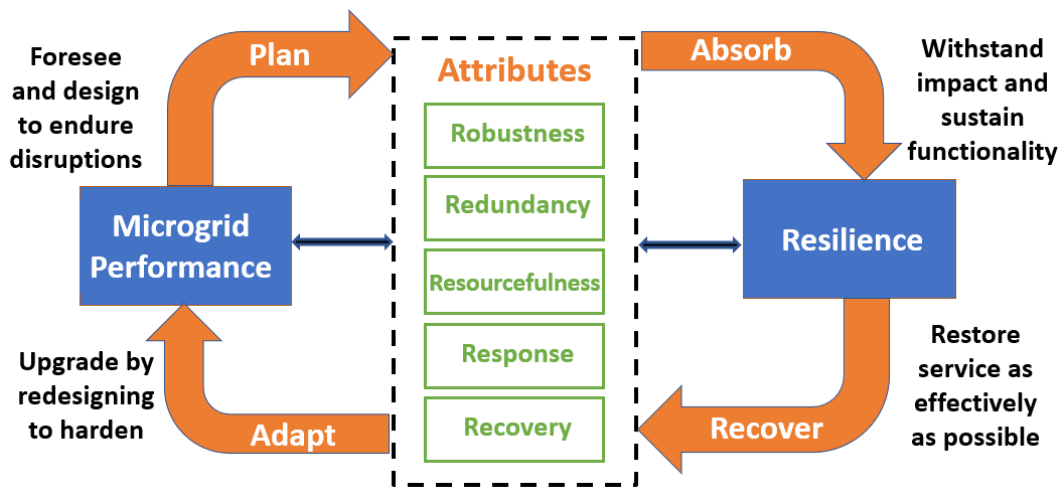

Figure 1 Process of building a resilient microgrid

Resilience can be longer-term adaptability to withstand or rapidly recover from unusual and extreme events. Unlike reliability, resilience is not just about restoring power (to more customers and in less time) after an outage. Resilience can focus more on critical power and leave noncritical customers off for longer durations to preserve limited local energy. Resilience is also focused on longer duration events where the outage impacts can grow exponentially over days and weeks. Thus, simple metrics like reliability, fail to capture the entire concept.

In the cyber dimension, it is also important to differentiate between cybersecurity and cyber-resilience with respect to energy systems operations. Cybersecurity refers to the process and methods for protecting the data (which could be voltage, frequency, price signals, frequency regulation signals, etc.), including clear information about where the data resides and how it is exchanged between various components of the microgrid and the utility grid [33] [34]. Cyber-resilience, on the other hand, refers to the ability of the energy system to maintain its operation in the event of a threat, and if impacted, the ability to quickly recover after the event has passed [28] [30]. Generally, cybersecurity's central focus is vulnerability reduction. On the other hand, cyber-resilience takes an integrated approach to minimize the risk, which is a combination of threat, vulnerability, and impact (details in Section III.C). The rest of this article delineates threats, vulnerability, and mitigation strategies for microgrid resilience—understanding and quantification of these three aspects lay the groundwork for defining an effective resilience metric for a microgrid.

III. METHODOLOGY FOR ASSESSING THREATS AND ASSOCIATED VULNERABILITIES

It is essential to enhance the ability of decision makers to assess and manage risks associated with microgrids that may change over time and vary geographically. Therefore, clear identification of threats and vulnerabilities and their likelihood and impact is the first step in building resilient microgrids. This section describes how to identify the highest risks to a microgrid by assessing threats and vulnerabilities specific the location or situation in which the microgrid will be installed.

*A. Threats*

Threats are anything that can damage, destroy, or disrupt utility grid or microgrid operation. In other words, threats are what we are trying to protect microgrid against. They are typically natural or human-induced hazards that are not within the site's control, such as wildfire, hurricane, cyberattacks, or physical attacks. Threats are identified through the review of climate data and state hazard assessments, and stakeholder interviews with site staff or emergency management teams from the surrounding community. Example microgrid threats to be considered are outlined in Table 1.

Table 1. Types of Threats

| Natural Hazards | Technological Hazards | Adversarial Hazards |
|---|---|---|
| Hurricanes | Grid outage | Inside bad actor |
| Flooding | Water- and wastewater- line disruption | Outside bad actor |
| Earthquakes | Pumping system failure on water- and wastewater- lines | Act of terror |
| Severe winter storms | Water damage to solar panels causing internal short-circuit | Cyber attack |
| Wildfire | Battery stored in extreme temperatures | Political upheaval |
| Hailstorms | Internal Combustion Engine based gensets located in areas with flammable vapors or gases | War |

*B. Vulnerabilities*

Vulnerabilities are weaknesses within the microgrid either in infrastructure or processes. Unlike threats, they are within the site's control and can be modified or mitigated to prevent or reduce the impact of a disruption. Vulnerabilities are identified through stakeholder interviews with energy managers, electrical engineers, maintenance staff, emergency managers, utilities, microgrid designers, operators, and end-users. They can also be identified through a review of contingency response plans and after-action reports following disasters or disruptions. Examples of microgrid vulnerabilities to be considered are outlined in Table 2.

Table 2. Types of Vulnerabilities

| Type of Vulnerabilities | Examples |
|---|---|
| Physical | • Lack of redundant backup systems<br>• Lack of accessible spare parts<br>• Single points of failure in electrical lines or generation sources. |
| Natural | • Equipment location prone to flooding, fire, winds, earthquakes and other natural disasters. |
| Cyber | • Lack of cybersecurity defenses<br>• Communication with external networks or the internet<br>• Data and communication leaks between information network, operational network, and other networks or the open internet. |
| Communication | • Single communications paths<br>• Dependence on digital networks<br>• Lack of redundant network components. |
| Human | • Lack of trained staff to operate the microgrid |

| | <ul><li>Inability of trained staff to access the site and the microgrid equipment during an emergency</li><li>Lack of written procedures or training for operating microgrid.</li></ul> |
|---|---|

Communications and information technologies are increasingly tightly integrated with the electric system and, in extension, with microgrids. Numerous advantages are offered by greater connectivity and automated communications in terms of reliable and efficient grid operations. These technologies have also increased customer participation by providing a new medium for microgrids and consumers to interact with the utility grid. On the flip side, they also expand the vulnerable cyber surfaces by offering additional vectors for intrusions and breaches. External sources of information exchange for a microgrid include weather forecasts, pricing signals, Internet-of-Things (IoT)-connected appliances (smart appliances which provide access to data/controls via the internet), frequency-regulation signals, peer-to-peer transactions through public networks, and so on. The more capabilities to intake and process various types of information from external sources is added, the more vulnerable the microgrid is to cyberthreats. This is because smart meters and other advanced communications technologies are exposed surfaces for attacks if vulnerable parts such as control interfaces, data buses, data communication channels, and remote debug ports are not protected well. IoT devices connected to a public network and electrical systems network simultaneously can be conduits through which cyberattacks can be executed.

### C. Risk and its Assessment

Assessing the risk essentially means finding a way to quantify the relative potential of damage that various threats in the environment can cause to the microgrid. Risk, therefore, is a function of threats exploiting vulnerabilities to impact the operations, and damage or destroy the assets. Thus, we define the risk factor metric based on the probability of a threat, the probability of a vulnerability's exploitation given that threat, and the impact of the vulnerability. This definition of risk factor has been extensively used in the literature [35] [36] [37] [references]. A microgrid's system-level risk assessment is the first step in building resilient microgrids.[2]

#### 1) Physical

Threats are scored based on their likelihood of occurrence using a low-medium-high qualitative scale, or a numerical scale (e.g., 1-10, with 1 being very unlikely and 10 being very likely). Natural hazard threats are typically scored using documented natural hazards, climate projections, and professional judgment. Non-natural hazard threats are scored based on information collected during stakeholder interviews. Vulnerabilities are scored based on probability of exploitation from a given threat, and the potential severity of their impact. Similar to threats, vulnerability scores can use a low-medium-high qualitative scale, or a numerical scale (e.g., 1-10, with 1 being low impact and 10 being high impact). Vulnerability probability scores should consider the likelihood that a system will be compromised if the threat is realized. This score is typically based on expert interviews with staff familiar with the site, such as energy managers, electrical engineers, maintenance staff, emergency managers, utilities, and microgrid operators. The score may be informed by data in after-action reports from past events and maintenance logs but is often based on expert judgment. For example, the probability of vulnerability of generation units, with respect to the threat of flood, would be higher for a site located in a coastal area with its assets placed on the ground-level, and lower for a site with elevated assets. Vulnerability impact scores should consider the geographic area of impact, the number of end-users affected, cost, safety, and environmental impacts. Vulnerability scores are based on the stakeholder interviews and document review and are inherently more subjective than threat scores because there is typically less data available to inform them.

To evaluate the relationship between threats and vulnerabilities, the threat likelihood score is multiplied by the vulnerability probability and impact scores to create a risk score for each specific threat-vulnerability combination. Risk is created when a threat can exploit an already present vulnerability in the microgrid. The magnitude of the risk is determined by the likelihood of the threat and vulnerability, as well as the scale of damage the vulnerability could cause if exploited. Therefore, risk factor is calculated as intersection of these three factors, as shown in **Error! Reference source not found.**. The risk score enables analysis and ranking of the risks to prioritize mitigation actions. Threats and hazards change over time, so the risk assessment should be updated on a regular basis.

---

[2] Risks for a microgrid can be assessed for both individual components, as well as at the system level. The risk associated with the failure of individual components in the microgrid, in lack of an external threat, will get categorized as a reliability concern. Whereas the system-level risk assessment, with respect to external threats, is a crucial step in gauging the resilience of the microgrid.

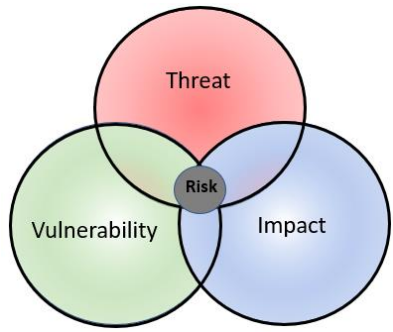

Figure 2 Risk Factor Metric - Venn Diagram

$$Risk\ factor \equiv (probability\ of\ threat)\ x\ (probability\ of\ vulnerability)\ x\ (impact\ of\ vulnerability) \qquad (1)$$

*2) Cyber*

Models for cyber-security threat analysis have been developed and studied by various technology companies' research teams. STRIDE [38] is one such model of threats developed by Microsoft for identifying computer security threats, which is commonly used. It is a mnemonic for the things that can go wrong in a cyber system, and stands for Spoofing, Tampering, Repudiation, Information Disclosure, Denial of Service, and Elevation of Privilege [39]. STRIDE model, however, doesn't yield a risk score as it is primarily focused on calculating threat scores.

The DREAD model scores a potential threat on five metrics (Damage Potential, Reproducibility, Exploitability, Affected Users, and Discoverability), and averages the scores to create numerical risk score. Table 3 provides the description and scoring method for the attributes of the DREAD model.

Table 3. DREAD model attribute and associated scores

| Category | Description | Scoring |
|---|---|---|
| Damage Potential | How much damage is can be done? | 0: No adverse consequences |
|  |  | 10: Total system failure |
| Reproducibility | How reliably does the attack work? | 0: Unlikely to succeed even with full access to the system |
|  |  | 10: Works consistently and requires at most the level of access available to the general public |
| Exploitability | How easy is it to perform the attack? | 0: N/A; by definition, every vulnerability must be exploitable in some way |
|  |  | 1: Difficult for even an expert on the affected system and vulnerability to exploit |
|  |  | 10: Minimal skill and no prior system access required |
| Affected Users | Scale of users affected | 0: No users affected |
|  |  | 5: Issue contained to a subset of users |
|  |  | 10: All users affected |
| Discoverability | How easy is it for an attacker to discover the vulnerability? | 0: Nearly impossible to discover, even with full source code and access to running system |
|  |  | 10: Readily visible to passive observer |
|  |  | NOTE: In practice, discoverability is assumed to be 10 for any specific vulnerability which is already known to the system owner/operator. This follows the logic that if you know about the vulnerability, it is likely that someone else does too [40]. |

The analogy between the physical threat model and cyber threat model can be established by loosely comparing the attributes as follows:
- probability of threat ≡ average of reproducibility + exploitability
- probability of vulnerability ≡ discoverability
- impact of vulnerability ≡ average of damage + affected users

## IV. Threats to Micro (and Macro) Grids and Associated Vulnerabilities

Threat modeling for complex systems is a four-stage process that begins with defining and diagramming the system, then identifying things that can go wrong, after which mitigation strategies can be developed. The entire process is validated to ensure accurate conclusions. For a complex cyber-physical system like a microgrid, there are multiple dimensions of resilience. They can be classified into three domains:

- *Physical resilience*: Sustaining physical infrastructure (which includes assets and electrical cables connecting the assets) during a disruptive event and continuing or restoring operations rapidly after the event is over.
- *Cyber resilience*: Identifying and defending against various types of cyberattacks and maintaining safe performance during the occurrence of such an event.
- *Communications resilience*: Maintaining safe and stable operational performance during communications failures from the utility grid end or internal microgrids communications.

The following subsections discuss these threats in more detail.

### A. Physical Threats

Physical threats include natural hazards (wildfires, hurricanes, floods), changing climate (more extreme temperatures), and human-induced attacks (including terrorist attacks, such as shooting substations).

- *Natural hazards*: Weather events like tropical cyclones, hurricanes, floods, earthquakes, and wildfires can cause damage to the utility grid, and if the epicenter of the impact is geographically close to the microgrid, then the impact can be of equivalent order. Examples of vulnerabilities resulting from natural hazards include:
    - Flooding can inundate equipment on or near the ground, cause generators to stop operating, and cause permanent damage;
    - Strong winds can cause microgrid components to fly away or topple overhead power lines;
    - Earthquakes can shatter the physical infrastructure of the microgrid, causing damage to assets;
    - Lightning strikes to electrical components or cables in the microgrid can burn the components; and
    - Wildfires can destroy generation, distribution, and controls equipment.
- *Changing climate*: increasingly extreme temperatures caused by the changing climate can put undue strain on a microgrid especially when it is operating in islanded mode. The spike in cooling and heating loads can cause harm in two potential ways: 1) excessive power demand can cause equipment and devices to operate above their rated capacity, resulting in permanent damage; and 2) increased demand can cause load-shedding in the microgrid. If load-shedding strategies are not automated, then there is an increased risk of a site-wide outage. Example vulnerabilities resulting from changing climate include:
    - Heatwaves can cause an abrupt rise in air-conditioning loads, and if generation cannot match the demand, site-wide outages or load-shedding can happen;
    - Winter storms and blizzards can bring extreme cold, snow, hail, ice, and high winds, and equipment overload due to prolonged periods of electric heating loads which can damage equipment permanently; and
    - During winter storms lasting several days, maintenance of the assets (damaged cables or generators located outdoors) becomes very difficult and increases the probability of a microgrid-wide outage.
- *Human-induced attacks*: Physical attacks, such as bombings or shootings of major microgrid components like generators and storage assets, could cause serious damage and disable a microgrid. Electromagnetic pulse attacks can also take out generation assets and controllers in the microgrid [41] [42]. The attacks can be from one gunman shooting a specific equipment to a massive bombing that can destroy the whole site.

### B. Cyber Threats

Cyberattacks on microgrids can be classified into the following three types [43] [44]: attacks on availability, integrity, and confidentiality.

- *Availability:* Attacks on availability attempt to limit or delay access to data. In a microgrid, this may be extended to include availability of power in addition to data. Attacks against availability are generally referred to as Denial of Service (DoS) attacks. DoS attacks in microgrids generally include not only blocking resource access but also delaying the timing of critical message exchange. For example, jamming communications channels between various parts of the microgrid.
- *Integrity:* Attacks on integrity interfere with the accuracy or trustworthiness of data. Spoofing, tampering, and elevation of privilege are all integrity violations. For the microgrid, interference with the quality of power might also be considered a violation of integrity. For example, an attacker uses a Domain Name System (DNS) hijack to feed inaccurate price data to the controller, showing diesel as cheaper than natural gas. The result is that the microgrid uses diesel instead of natural gas, costing the operator money. Faking a microgrid disconnect from the grid signal compromises the integrity. Other examples of integrity attacks include sending false commands to the microgrid controller, causing unwanted or premature islanding; and changing the setting groups of protection relays in a microgrid, causing the relays to trip erroneously.

- *Confidentiality:* Attacks on confidentiality are attempts to gain access to privileged information. This might include operational or billing data, or any other information that should not be accessible to unauthorized entities. For microgrids participating in markets, the bidding price data can be stolen.

The example microgrid (shown in Figure 3) is a cyber-physical system containing physical infrastructure (generation, storage, local distribution and transmission, loads) equipped with electronic controls and sensors as illustrated. These electronic devices are then connected via a network to a supervisory control system, which monitors all systems in real-time and coordinates the operation of the various physical systems. This supervisory control system receives input from and reports back to the microgrid control system, which is generally a near real-time software system that performs optimization and resource scheduling. The microgrid control system is also connected to the utility, any other external data sources, and the human-machine interface (HMI).

In this system, each interface or trust boundary represents a potential attack surface that must be analyzed for vulnerabilities. As methods for identifying specific threats are extensively well-documented elsewhere, we will focus here on those attack surfaces that are common to microgrid systems in general, rather than on specific attacks, which must necessarily be identified on a case-by-case basis. Table 4 summarizes the vulnerable attacks surfaces along with the types of threats that can impact the operations of microgrids.

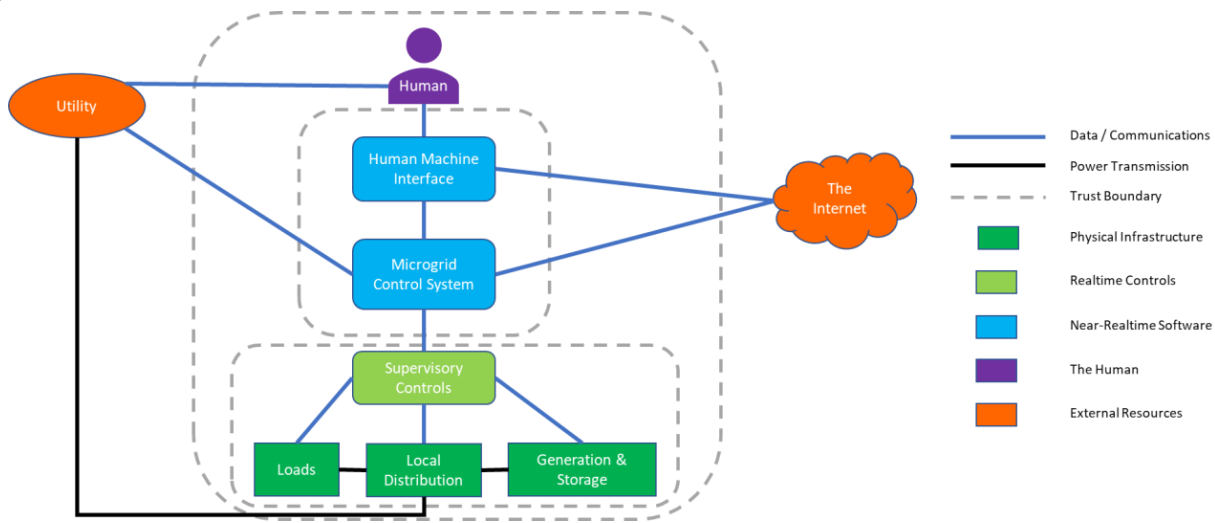

Figure 3. Microgrid diagram with cyber-physical layers and associated trust boundaries

Table 4. Cybersecurity Risk Analysis

| Attack Surface | Examples | Threats/Vulnerabilities | Potential Impact |
| --- | --- | --- | --- |
| Wired Links | Ethernet, Phone lines, RS-232, RS-422, RS-485, I2C (often used for sensors) | Line may be cut or unplugged, causing DoS. An attacker may splice into the wires and eavesdrop, tamper with, and inject data into the medium. | Blocking resources Delaying the timing of critical message exchange |
| Wireless Links | IEEE 802.11 (WiFi), Cellular networks, ZigBee/XBee | The broadcast nature of the medium enables anyone to listen to communications. Signals may be jammed, resulting in DoS. | Confidentiality lost With jammed signals, SCADA system loses connection with microgrid's analysis layer, causing sub-optimal operation. |
| Unencrypted Communications | Most low-level serial protocols including I2C and SPI Unencapsulated TCP and UDP, including HTTP, FTP, Telnet | An attacker who can access the medium can view, modify, and send meaningful data. | Data confidentiality lost—market participation bidding strategy compromised. |
| Unauthenticated Communications | DNS, HTTP, most low-level serial protocols | Attacker can pretend to be legitimate data source or user. | Can cause unwanted islanding of microgrid or unwanted change to grid-connected mode during island (can be a life-threating issue if |

| | | | maintenance workers are working on electrical components). |
|---|---|---|---|
| Exposed Endpoints | Sensors, HMI, Remote Servers, Anything the attacker can gain physical or virtual access to | Attacker could falsify actionable data. Attacker could damage interface, causing DoS. | Wrong price-signal or weather data will cause suboptimal and noneconomic operational dispatch strategy of the microgrid. |
| Human | Technicians, administrators, contractors, any user with the system or network access | Attacker gains real credentials from a legitimate user and can then access normally restricted systems. Insider threat – a legitimate user exploits their access to cause damage | Unwanted islanding Dispatch strategy causes economics losses in operations. Unwanted outages |

For microgrids with advanced controls and analytics capabilities, maintaining cyber-reliability is a matter of concern as well. The more complex the control algorithm, the higher the risk associated with having unforeseen/blinding scenarios making it behave in an unintended manner. Moreover, the higher the sophistication of the software stack, the greater the risk of injecting unintended bugs while developing the stack further. Since these challenges are not imposed by an external entity, they are not categorized as threats against which resilience solutions should be built in the microgrid. But they do pose a challenge for maintaining the reliability of microgrid, which is a significant aspect as well.

### C. Communications Failure Threats

The modern grid is envisioned to be highly integrated with multidirectional data exchange and information communication between all its parts. The vision is consistent with microgrids as well for deploying a more active and responsive infrastructure for faster demand response and more exhaustive energy management capabilities. It is indispensable for an intelligent microgrid to require a dependence on communication networks. On the other hand, using sophisticated communication schema makes the microgrid vulnerable, due to the increased risk of communication failure, along with the microgrid's growing dependence on electric systems and public communication networks.

An example of the severity of impact that a communication failure can have is 2013 Northeast blackout in the United States and Canada. It was primarily caused by communication failure when a small transmission disturbance failed to send out the alarm, making the operators unaware of the need to redistribute power flow. This initial trigger could have been controlled without causing a widespread outage; however, because of the delay in reacting to the overloaded transmission lines, a manageable local blackout cascaded into grid failure in the entire Northeast, leaving most of the region without access to power for two days to a week [45].

In the case of microgrids, if the communication failure happens during grid-connected operation, the microgrid's controller will lose access to grid-side information, resulting in its inability to maintain the commanded power and power-factor at the point-of-interconnection. The microgrid should be able to maintain stable operations and stop feeding the power to the grid if it has been doing demand management before the communication failure. In worst case scenarios, the microgrid needs to have the ability to island itself safely. Communication failure internal to the microgrid is more of a reliability concern, as it is not imposed by an external entity. If not addressed appropriately, it can adversely impact the operation of the microgrid and damage the assets if left unattended for longer durations especially when operating in islanded mode.

### D. Threats Due to Interdependencies Between Various Systems

Electricity, oil, natural gas, transportation, telecommunications, and water systems are all interdependent critical infrastructure (CI) systems. The operations of power systems are in constant exchange with other CI systems. On one level, electric power systems support other CI systems in their operations, whereas, on the other level, electric power systems depend on other CI systems to be able to operate. The interdependencies between all CI sectors is thoroughly mapped in [6]. Power systems provide electricity to:

- Natural gas systems for operating their control systems and storage
- Oil industry for pumping, control systems, and lift stations
- Telecom systems for switches and signaling
- Water infrastructure for purification and pumping.

On the reverse side, other CI systems provide the following material and services to facilitate utility grid and microgrid operations:

- Natural gas and oil industries provide fuel for generators;
- Transportation industry transports operators and sometimes diesel fuel to power plants and generators;
- Telecom systems provide a communication layer and SCADA systems for cyber layers; and
- Water is required for construction and infrastructure recovery of power plants, as well as operations including cooling, productions, and so on [46].

These tightly intertwined interdependencies between other CI systems and electric power systems increase their vulnerabilities by exposing them to second-order and third-order threats. Microgrids are also equally, if not more, vulnerable to the threats posed by such interdependencies. Microgrids operate as integrated energy systems where gas, water, distributed generation, and utility grid supplies are coordinated to supply various kinds of loads, as depicted in Figure 4. The electric energy supply (shown in green) is partly dependent on the electric energy supplied from the CHP unit; the CHP unit, in turn, relies on natural gas and water inputs. Similarly, to serve the heating load, the boiler needs both water and natural gas. Supply of the water load requires both electricity (to operate the pump) and water itself. Therefore, interruption in the supply of any of these critical utilities (electricity, natural gas, water, or heat) can negatively impact the ability of the integrated energy system to meet loads and affect the operational economy of the microgrid.

The threats, vulnerabilities, and potential impacts of different CI system failures on the microgrid are described in Table 5. The risk score methodology for this kind of threat is similar to physical threat risk score calculation method described in section III.C.1), where the risk factor is a multiplication of the probability of threat, probability of vulnerability, and impact of vulnerability.

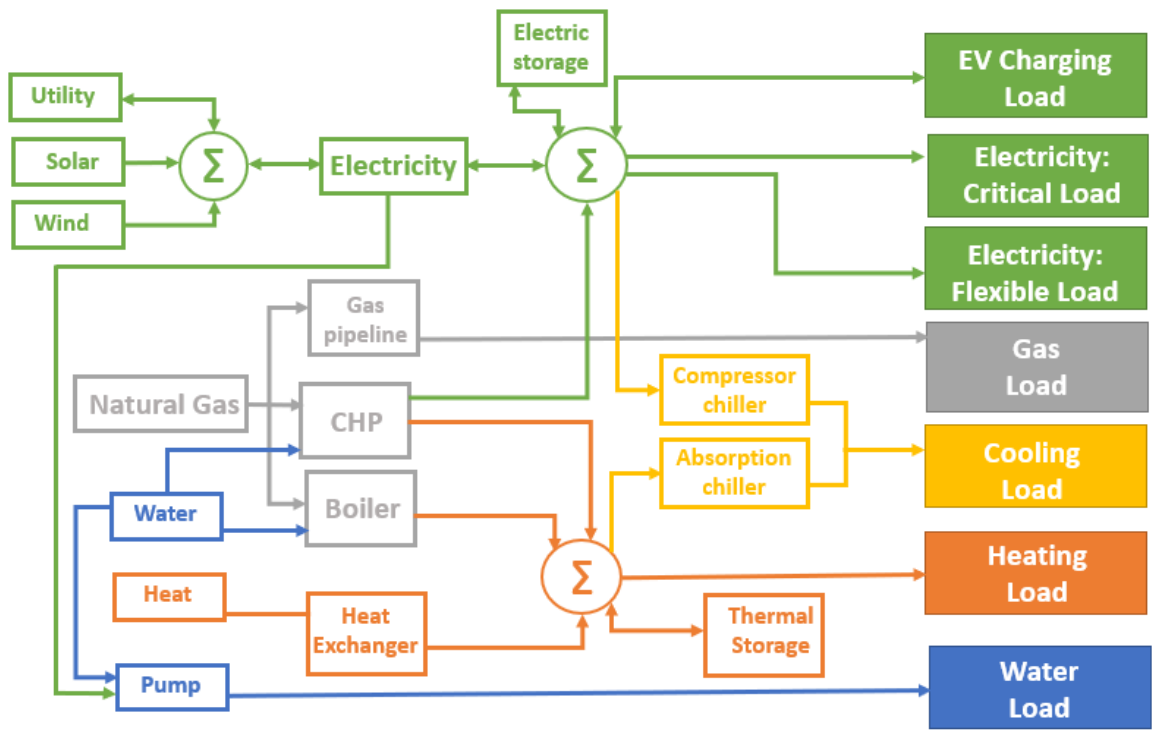

Figure 4 Microgrid operation as an integrated energy system[3]

Table 5. CI Systems Interdependencies Risk Analysis

| Threat | Vulnerability | Potential Impact |
|---|---|---|
| Natural gas supply interruption | No input fuel for heating system | -Heating load cannot be served<br>-If microgrid also has electric heating system paired with gas heating then undue load on electric heating system, which may cause component failures. |
| Water supply interruption | CHP and boiler need water, the operation of which can be halted | -Not enough energy production to supply electric and heating load |
| Utility grid outage | No external electricity supply source | -If microgrid is not designed to sustain a 100% load in isolated-operational mode, then unintended load shedding.<br>- If microgrid is not designed for bump-less transfer or black starts, then powering it up without grid-supply is not possible |

---

[3] Diagram adapted from [6].

## V. CASE STUDY: EXAMPLE RISK SCORE CALCULATION

In this section, we first introduce an example microgrid. The physical and cyber risk factors are then calculated in subsections A and B. Since a microgrid is inherently a cyber-physical system [47] [48], it is important to integrate the risk assessments of these two layers to holistically evaluate microgrid resilience. The combined risk assessment results provide an effective way to prioritize which mitigation measures will have maximum impact on increasing the resilience of the microgrid.

### A. Physical

To illustrate an example of quantifying a risk factor, Table 6 lists the three attributes of risk factor calculation. The test microgrid is assumed to be located on Florida's coast, and a hurricane is the threat being assessed. The probability of a hurricane hitting the location is 90%, based on historic and projected climate data. If a hurricane does hit the site, there is a 70% probability that the generators (including ground-mounted solar and electric battery storage) will be flooded, based on expected storm surge levels at the generator locations. If the generators are flooded, the expected impact is severe (scored a 9 out of 10), because the flooded generators will not function and the microgrid will not be able to serve critical loads. The total risk score is 9 x 7 x 9 = 567. This is a high-risk scenario, as both the likelihood of the hurricane and the potential damage is high.

**Table 6. Physical Threat and Vulnerability Scores**

| Category | Score | Reason |
|---|---|---|
| Probability of threat | 9 | 90% likelihood of hurricane occurring in Florida |
| Probability of vulnerability occurring (given the threat) | 7 | 70% probability that generators (solar panels and battery) will be flooded with muddy waters if hurricane occurs |
| Impact of vulnerability | 9 | 90% impact (because flooded generators will almost certainly cause the microgrid to no longer be able to serve critical loads as well as the generator themselves will be damaged and will need repair/replacement based on the degree of damage) |

Risk Score: 9 x 7 x 9 = 567

### B. Cyber

To illustrate one method of quantifying the severity of a vulnerability, we will look at a (simplified) example system shown in Figure 5. The system consists of a single inverter-based generation unit that delivers power to several loads. The loads contain sensors, which report input voltage and frequency information back to the supervisory controller, which in turn provides feedback to the inverter controls of the generation unit so that it can adjust its output to ensure the correct voltage and frequency arrive at the loads. For this scenario, we assume the sensors communicate to the controller via an unencrypted ZigBee link.[4] Using the DREAD model (explained in section III.C.2), we can quantify the severity of various attacks against this system. The vulnerability we will specifically examine is the unencrypted wireless link.

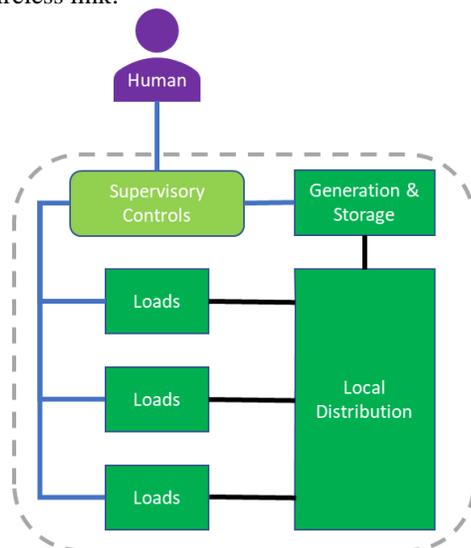

**Figure 5. Test microgrid for cyber risk assessment example**

For the first attack (Table 7), the attacker is assumed to be passively observing the network. Because voltage and frequency in electric grids are standardized, it would be simple for an eavesdropper to guess the nature of the communications. Having done this, the eavesdropper can then extract the detailed frequency and voltage information, and learn the exact data format being used (which is not necessarily damaging by itself but enables attack number three below). Eavesdropping of this nature does require some equipment, but all the required equipment is readily available at low cost. The data gathering itself can be performed by an unauthenticated user, but this is partially offset by the significant knowledge required to analyze the data. If the attacker is assumed to be placed near the supervisory controller's receiver, then all loads' data is compromised.

---
[4] This example system is deliberately unsecure to make scoring simple.

Table 7. Cyber Threat Score: Data-Sniffing Attack (Attack 1—Sniffing)

| Category | Score | Reason |
|---|---|---|
| Damage | 2 | The attacker receives detailed information about voltage and frequency, as well as the data format. Some usage information could be extrapolated. No direct harm is done to infrastructure. |
| Reproducibility | 5 | Some tools required, but all required tools are readily available at low cost. |
| Exploitability | 5 | Can be done by an unauthenticated user, but requires significant knowledge to use results. |
| Affected Users | 10 | All loads' power consumption compromised. |
| Discoverability | 10 | By convention, discoverability is normally assumed to be 10 for any critical system, based on the assumption that if you can discover the vulnerability, so can someone else. |

Risk Score: 2+5+5+10+10 / 5 = 6.4

This is a moderate threat, but since the damage potential is very low and the knowledge required to perform the attack is high, this is unlikely to be a high priority.

In the second attack (Table 8), the attacker performs a DoS attack against the wireless network by jamming the supervisory controller's receiver. This causes a complete loss of visibility for the controller and reduces the ability of the controller to handle unforeseen events, but does not directly cause any permanent damage to the system (It is assumed for this example that the feedback provided by the loads is not critical to system stability under normal conditions.). This attack can be carried out very reliably but requires either specialized hardware or a readily available device specifically programmed for the task. All technical knowledge required to perform this attack is available on the internet, and no authentication is required. Exploitability is downgraded from 10 to 7 because of the need to place hardware near the target. Assuming the stability of the system is affected at all, all loads are affected. With jamming in particular, the vulnerability of wireless systems is widely known—even appearing in internet comics [49].

Table 8. Cyber Threat Score: DoS Attack (Attack 2—Signal Jamming)

| Category | Score | Reason |
|---|---|---|
| Damage | 7 | Complete loss of visibility, but no (direct) permanent damage (assuming system can still operate in an open loop configuration). |
| Reproducibility | 5 | A compatible device on the same channel can trivially and reliably inject garbage data. |
| Exploitability | 7 | Minimal knowledge required, no authentication. |
| Affected Users | 10 | All loads' power supply stability decreased. |
| Discoverability | 10 | By convention, discoverability is normally assumed to be 10 for any critical system, based on the assumption that if you can discover the vulnerability, so can someone else. |

Risk Score: 7+5+7+10+10 / 5 = 7.8

This is a much more severe threat than just eavesdropping. The attack is relatively easy to carry out and requires minimal knowledge of the actual system. It also has a much higher impact.

For Attack 3 (Table 9), the attacker is assumed to have already performed Attack 1 and has detailed knowledge of the loads and the general stability of the network. By sending inaccurate frequency and voltage feedback to the controller, the attacker can cause the controller to change its behavior in such a way as to damage loads or cause the generator to shut down. This attack is fairly well understood but requires significant and complex tooling to function beyond a simple denial of service attack. While no authentication is required to perform the attack, advanced skills and specific knowledge of the exact system are required to accomplish the attack. If successful, the attack affects all users of the system, as well as the system itself. The vulnerability of the system is assumed to be known, so discoverability is assumed to be 10.

Table 9. Cyber Threat Score: Tampering, Data Injection Attack (Attack 3—Incorrect Frequency/Voltage Data)

| Category | Score | Reason |
|---|---|---|
| Damage | 10 | Incorrect frequency or voltage can cause total shutdown or permanently damage loads. |
| Reproducibility | 3 | This attack is well-understood but requires significant and complex tooling to function beyond a simple DoS attack. |
| Exploitability | 3 | No authentication required, but advanced skills required to accomplish. |
| Affected Users | 10 | All loads and generation node potentially damaged. |
| Discoverability | 10 | By convention, discoverability is normally assumed to be 10 for any known vulnerability in a critical system, based on the assumption that if you can discover the vulnerability, so can someone else. |

Risk Score: 10+3+3+10+10 / 5 = 7.2

This threat is also moderate to high severity because, while it potentially has catastrophic effects, it has a low probability of functioning, and requires significant resources to accomplish.

Since all three of these threats target the same vulnerability—the unencrypted wireless link—we assign the highest score to the discoverability attribute (which is equivalent to probability of vulnerability attribute, as explained in the end of Section III.C.2). Figure 6 shows a radar plot of the three threats and their corresponding five attribute scores based on the DREAD model.

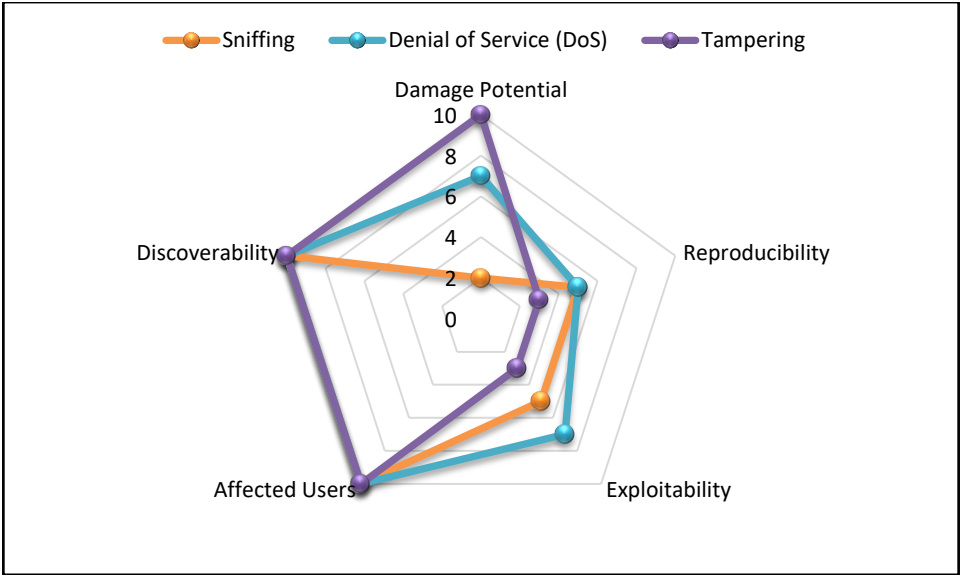

**Figure 6 DREAD Scores for 3 Cyber Threat Examples**

### C. Overall Risk Factor Calculation: Cyber-Physical System

To conduct a comprehensive risk analysis for making design decisions geared toward resilience, this section presents an overall risk factor calculation for the microgrid in our use-case (a detailed description can be found in sub-sections A and B). Table 10 shows example risk scores (function of probability of threat, probability of vulnerability, and impact of vulnerability) for various combinations of input threats and vulnerabilities for the test microgrid assumed in sub-section A. For the cyber threats, the three attributes of the risk score are calculated based on the analogy between the DREAD model and risk assessment methodology for physical threats, which is covered in section III.C.

**Table 10. Example Risk Score Calculation for Example Test Site[5]**

| Threat | Type of Threat | Vulnerability | Probability of Threat | Probability of Vulnerability | Impact of Vulnerability | Risk Score |
|---|---|---|---|---|---|---|
| Higher storm surge due to hurricanes (storm) | Physical | Generators, solar panels, and battery are at ground level in areas prone to flooding | 9 | 7 | 9 | 567 |
| Increased number of days with thunderstorms/lightning (lightning) | Physical | Lack of accessible spare parts | 8 | 8 | 7 | 448 |
| Increase in coastal land loss | Physical | Critical microgrid infrastructure near coast | 7 | 5 | 5 | 175 |

---

[5] These numbers are approximate and are only intended for showing examples, not to be used as risk factors for actual microgrid design.

| | | | | | | |
|---|---|---|---|---|---|---|
| (land-loss) | | | | | | |
| Increase in magnitude of hottest annual temperature (heatwave) | Physical | Lack of redundant backup systems | 6 | 4 | 6 | 144 |
| Increase in the number of tornados (tornados) | Physical | Equipment is outside in an unprotected area | 4 | 3 | 2 | 24 |
| Earthquake | Physical | Lack of staff trained to operate the microgrid | 2 | 4 | 8 | 64 |
| Data sniffing | Cyber | Unencrypted wireless connection | 5 | 10 | 6 | 300 |
| DoS | Cyber | Unencrypted wireless connection | 6 | 10 | 8 | 480 |
| Data Tampering | Cyber | Unencrypted/unauthenticated wireless connection | 3 | 10 | 10 | 300 |

Figure 7 shows the risk scores in a 3D bar plot where the threat probability is depicted by the colors of the bars (also shown in upper left legend). The data set is four-dimensional where the risk score (the dependent variable) is a function of the probability of the threat, probability of vulnerability, and impact of vulnerability (three independent variables). To visualize the information in three dimensions, the probability of threat variable is qualitatively shown by the colors of bars; however, the plot is effectively showing the relative difference between the risk score for different kind of physical and cyber threats.

Based on the calculated risk factors for different kinds of threats and their impact on the microgrid, the mitigation strategy should prioritize the flooding protection (increasing elevation of generators) and storm preparedness (keeping inventory of spare parts) measures over the seismic design to make it resilient against hurricanes. Also, for this test microgrid system, the cyber risks are relatively higher. Therefore, the mitigation strategy should also make it a priority to build software features capable of defending against such threats.

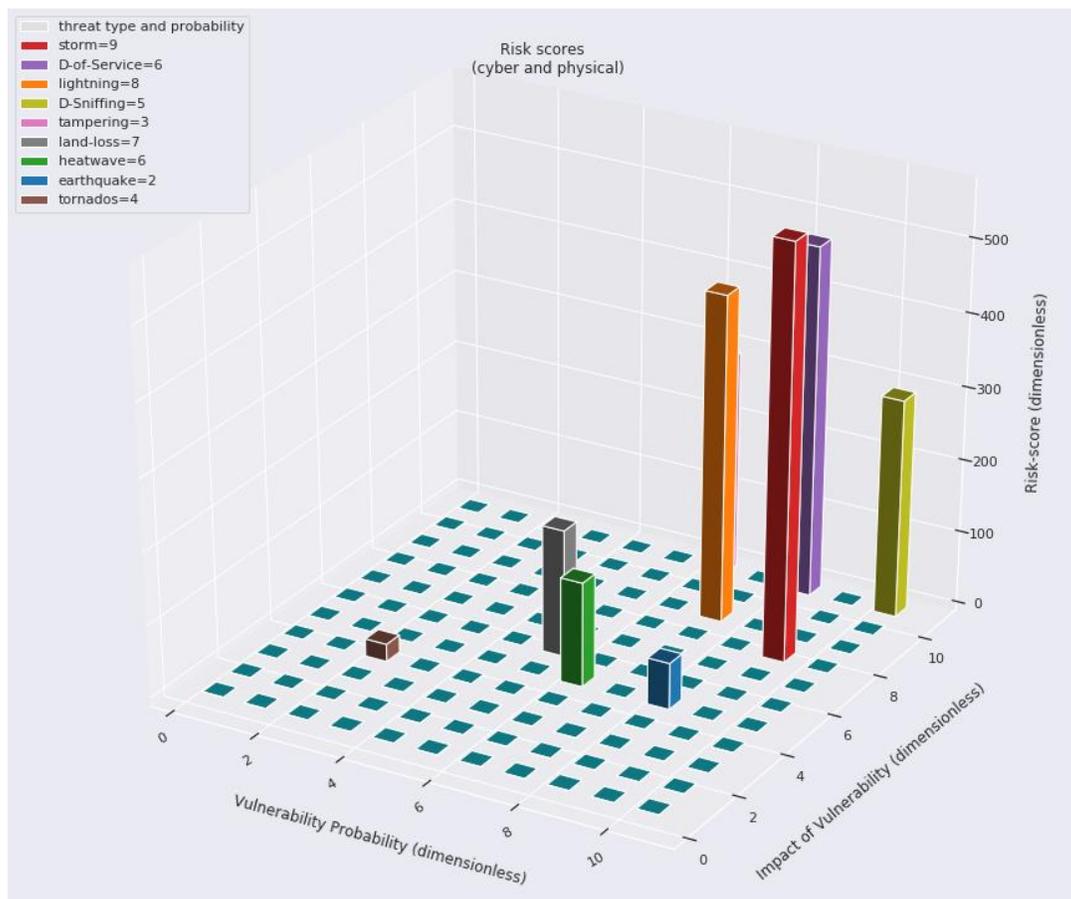

**Figure 7 Risk scores for various kind of threats to example microgrid**

## VI. MITIGATION STRATEGIES

After understanding the risks, the next step is to identify mitigation options that reduce the probability or impact of each vulnerability to respective threats. Each mitigation strategy is then evaluated based on its potential to reduce the risks against its complexity and cost of its adoption. When developing mitigation measures, actions and measures that reduce risk in a range of ways should be considered. Table 11 delineates a useful way of assessing the strength of resilient microgrids through five attributes (also called 5R) and provides examples of how these attributes increase resilience in different ways. This 5R approach of classification does not encompass various phases of resilience modeling, as depicted in Figure 1, nor does it provide precise actionable mitigation strategies targeted for the pre-disaster and post-disaster phases' operations and recovery, as depicted in Figure 8.

Table 11. Resilience Characteristics and Example Mitigations[6]

| Attribute | Qualities | Examples |
|---|---|---|
| Robustness | <ul><li>Physically secure</li><li>Cyber secure</li><li>Hardened infrastructure</li><li>Performance monitoring</li></ul> | <ul><li>Risk Management Framework (RMF)[7] -compliant control systems</li><li>Active vs. passive performance monitoring</li><li>Maintenance schedule and checklist</li><li>Physically enclose microgrid equipment inside mechanical rooms to protect from elements and unauthorized persons;</li><li>Seismic design in earthquake zones; elevated platforms in floodplains</li><li>Cyber-secure access to controls and networks.</li></ul> |
| Redundancy | <ul><li>Eliminate single points of failure</li></ul> | <ul><li>Modular units of accounting for maintenance and downtime</li><li>Redundant lines (power and comm) and equipment</li><li>Backup staff (microgrid operator).</li></ul> |
| Resourcefulness | <ul><li>Available power generation</li><li>Energy storage</li></ul> | <ul><li>Diversified generation sources including generators, renewable energy, and storage;</li><li>Load shedding to prioritize more critical loads</li><li>Uninterruptable power supply (UPS).</li></ul> |
| Response | <ul><li>Automated</li><li>Self-healing</li><li>Forecasting/threat assessment</li><li>Performance indicators</li><li>Training and exercises</li></ul> | <ul><li>Maintenance staff training and exercise</li><li>Data collection and predictive analytics</li><li>Fault tolerance (failover or failsafe)</li><li>Inclement weather response plans</li><li>Smart control systems</li><li>Documented procedures available during an emergency.</li></ul> |
| Recovery | <ul><li>Standardized components</li><li>Spare parts inventory</li><li>Damage Assessment</li><li>Prioritization of re-powering</li></ul> | <ul><li>Maintain spare parts inventory, preferably using commercial off-the-shelf parts;</li><li>Utility coordination and agreements</li><li>Black start sequence.</li></ul> |

---

[6] Adapted from Air Force Civil Engineer Center.
[7] Risk Management Framework (RMF) [50]

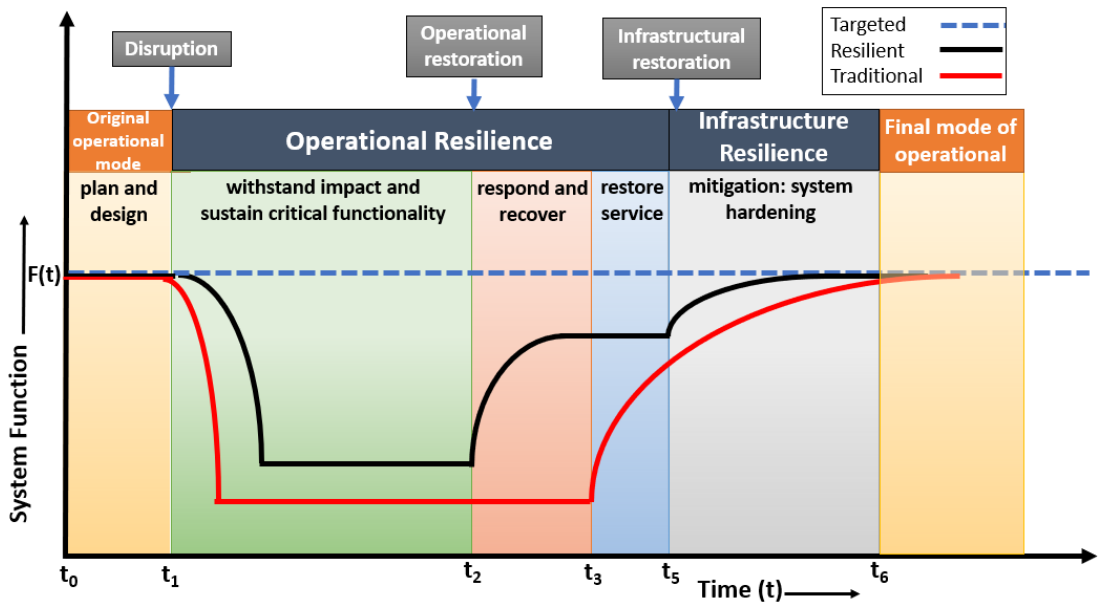

Figure 8. Comparison of resilient and traditional microgrid operations through disruptive event [8]

Therefore, in this work, we propose a comprehensive mitigation strategy encompassing both infrastructural and operational measures to make the microgrid resilient during pre-, during-, and post-disaster phases. As illustrated in Figure 9, mitigation strategies can be broadly classified into pre-disaster planning, during-disaster measures, and post-disaster recovery and restoration. The pre-disaster planning phase includes assessment of threats, associated vulnerability, and implementation of system-hardening measures (both physical hardening as well as cyber strengthening) to minimize the probability of damage to the microgrid. Preparation can also include restocking supplies, relocating vehicles, etc. just prior to extreme weather or based on high threat level. Actions during-disaster could include deliberate de-energization, isolation of damaged lines, and status checks of essential staff (recall alternate individuals if primary is unavailable). Post-disaster recovery mitigation strategies focus on speedy recovery actions and restoration of energy supply. The following two sections discuss these mitigation measures in detail.

---

[8] Diagram adapted from [6].

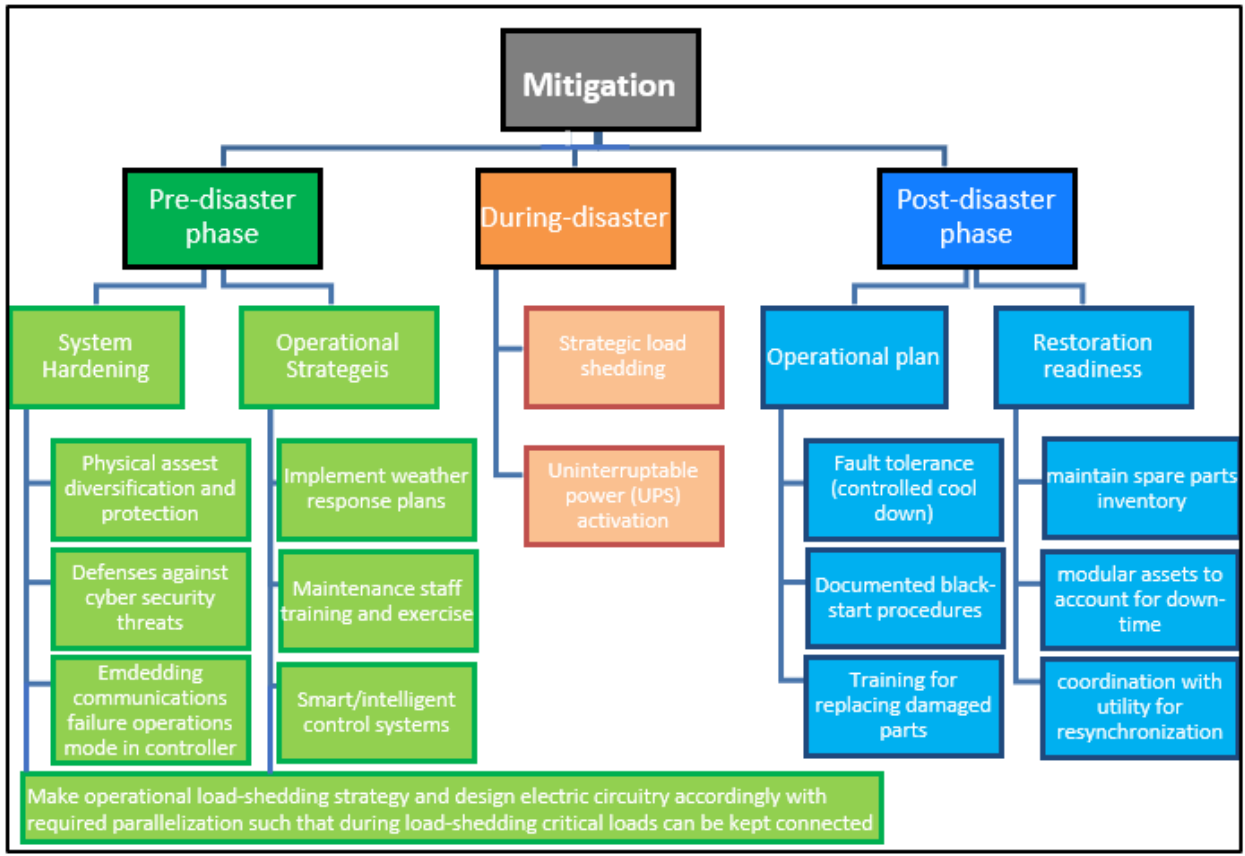

**Figure 9. Infrastructure and operational mitigation strategies for a resilient microgrid**

### A. Designing and Building Resilient Microgrids

In the planning phase of microgrid design (depicted by the Original Operational Mode section of Figure 8), a set of mitigation measures must be considered for making microgrids resilient during a disruptive event (i.e., increasing microgrids' robustness or resistance and maintaining supply). Therefore, these mitigation measures are aimed at minimizing the damage and attempting to keep the supply (at least the critical loads) uninterrupted during the event. The measures primarily make physical assets and controls software less vulnerable to attacks and are categorized as system-hardening mitigation options, whereas day-to-day operational procedures and on-site staff's competence in operating the microgrid in response to disruptive events come under operational strategies. Coordination of both System Hardening and Operational Strategies is required for the measure where strategic load shedding capabilities are designed and constructed. The electric circuitry of the microgrid grid should consider modular and/or parallel design for enabling strategic-load shedding.

#### 1) System Hardening

System-hardening methodologies are as follows:
- Diversifying the generation sources, including generators, renewable energy, and storage. For microgrids located on islands, diversification of generation resources with renewable energy also may decrease the cost of operation; generator fuel can be costly in remote places, primarily due to cost and risk associated with its transportation over long distances. For microgrids serving critical services, such as defense bases and financial institutions, incorporating a UPS is another way of increasing resourcefulness.
- Protecting assets from severe weather disruptions by implementing system-hardening measures, such as bolting down the equipment (to protect from strong winds), elevating the equipment above the floodplain, or enclosing equipment inside mechanical rooms (to protect from bad actor or other disruptive elements).
- Designing critical electrical circuits to avoid any single point of failure and modularizing the connections to account for the maintenance and downtime of the assets.
- For a certain category of threats (i.e., unusual fault current from the utility grid side), the component should be designed to operate under a range of current and voltage conditions.
- Installing a control system that is RMF compliant [50] can harden or strengthen the microgrid to improve information security and resilience against cyberattacks.

- The cyber layer of the microgrid can be made more resilient to DoS and loss of communication by increasing the sophistication of the control systems and analytics software implemented on top of the control systems.
    - In the event of loss of communication with the utility, switch to island mode and alert user;
    - In the event of total communication loss (i.e., between microgrid control system and SCADA), all nodes should have a default behavior.
- The cyber layer can be hardened against DoS and intrusion by adding a firewall between the microgrid control system and the internet
- Looking back at the threat assessment examples presented previously, each attack surface can be hardened to reduce the ease of exploitation.
    - Wired connections can be buried or otherwise made inaccessible to attackers;
    - Wireless links can be made directional and support multiple bands to reduce the impact of jamming and interference;
    - Redundant physical connections (wired, wireless, or both) reduces chances of total loss of communication;
    - All communications can be encrypted and signed to mitigate eavesdropping, tampering, and spoofing;
    - Restricting access to physical infrastructure reduces attack surface; and
    - Training users to follow strong operational security practices reduces opportunities for attackers to obtain legitimate credentials.

### *2) Operational Strategies*

Day-to-day operational strategies that can make microgrids more resilient include:
- Performance monitoring of the microgrid assets (both hardware and software platforms) for preventive maintenance of the assets.
- Predictive and condition-based maintenance using data analytics and predictive modeling can help, as this process helps with increasing the reliability of components; fault-prone equipment is flagged with predictive maintenance and can be repaired or replaced;
- A maintenance schedule (with checklist) must be followed; this schedule can be assisted by performance monitoring and predictive maintenance software;
- Maintenance staff must be trained and prepared to respond to extreme events and be able to operate the microgrid in manual mode if the need arises;
- When weather conditions deteriorate (or extreme weather alerts are released), response plans must be in place that can shut down the vulnerable elements and yet sustain critical loads (e.g., with the help of UPS); and
- Documenting procedures for maintenance and response during the disruptive event is very important. Also, equally important is the easy accessibility of the documents (for example, hard copies instead of an electronic copy).

### *B. Response During-Disaster Event*

These mitigation strategies apply to the section depicted by the Operational Resilience section of Figure 8. For enabling the microgrid to operate during a disruptive event (for the cases when the disruptive even is predictable, such as windstorm alert or flood warning) to serve critical loads (at the minimum), the following operational steps can be taken:
- Demand-side resources: load-shedding strategy in place that lets the controller turn off the non-critical loads during the event and switch on the UPS resources;
- Switch UPS on: switching UPS on should be a semi-automated process that kicks off as soon as demand-side measures are activated;
- Generation rescheduling: for grid-connected microgrids, the on-site generation can be rescheduled to dispatch at a later point in time, as the utility grid outage is a high-probable threat after the disaster has hit;
- Network reconfiguration: if the electric circuitry design allows, the network can be reconfigured to connect generation and storage with critical loads only (as opposed to normal operations, where all critical and non-critical loads are connected), for avoiding the situation where the radial nature of cables running through critical and non-critical cause the critical load to be shed because of a fault in the non-critical load pathway; and
- Optimal parameter settings of droop-controlled units (i.e., generators): the parameters should be set to support conservative voltage regulation to help survive the event. In other words, the relaxed parameter limits will allow low-amplitude voltage fluctuations without tripped the microgrid.

### *C. Post-Disaster Readiness*

Once the disruptive event has passed, the characteristic of a resilient system is to be able to respond to the damage proactively and recover faster. This requires the site to have sufficient redundancy, as well as operational plans for rapid restoration.

### *1) System Restoration*

For rapid system restoration, the following measures can be taken:

- Maintain a spare parts inventory;
- As much as possible, use commercial off-the-shelf parts and equipment for constructing the microgrid so that buying spare parts is not a challenge; and
- Follow modularized circuit design such that a subset of assets can be brought online without replacing all damaged parts.

   *2) Operational plan*

- Staff must be trained for executing black starts, and the process should also be documented well;
- In the case of electric system faults, the assets must be equipped with a process for controlled cool-down for safe recovery; and
- For grid-connected microgrids, the resynchronization of the microgrid with the utility grid must be coordinated.

### D. Deployment Considerations

Though a range of mitigation strategies are available and technically feasible, there are some challenges that make it difficult to justify the capital investment in implementing these strategies. Moreover, inadequate information-sharing processes between utility and microgrid owners can hamper the microgrid's cyber layer from defending and mitigating certain types of cyber threats, due to lack of visibility into the utility's state.

In real-life microgrid deployment scenarios, resources and budget are largely, if not always, constrained. In such cases, it is very important to prioritize the mitigation measures presented in this section. The criterion of prioritization includes: 1) risk reduction capability; 2) difficulty of adaption and/or deployment; and 3) capital investment cost. As an example, if a microgrid is located in a woodland in a geographically dry area (e.g., California) and is installed for a military base, then the most effective mitigation strategies would be:

- Physical hardening: keeping the key assets, including generators, fuel-tanks, and batteries, in a mechanical room (to prevent direct shooting), which has its walls coated by a fire-resistant material (to prevent wildfires from setting assets, including a flammable liquid tank, on fire). Seismic design of the mechanical room for earthquake prevention.
- Cyber hardening: being a highly susceptible location, the military microgrid's cyber layer should be hardened against confidentiality and integrity attacks by building software stacks equipped to defend against such attacks, or, at the very least, identify the attacks before larger damage is done. The hardwired communication lines should be grounded with water resistant coverings.

In the microgrid design and deployment process, after the risk-analysis, identification and prioritization of mitigation strategies, then the next steps are: 1) development of an action plan; 2) implementation or deployment of the action plan; 3) validation (i.e., not waiting to test the effectiveness of mitigation measures for the first time when a disaster hits); and 4) re-assessing the plan based on the validation test results. As described in Section II.B, building resilient microgrids is an iterative process (Figure 1). The above four steps must be continually re-assessed periodically to ensure the resilience of the microgrid with changing external circumstances.

## VII. Summary and Outlook

As the risks associated natural and human-induced threats continue to grow, power system resilience has become an important requirement for ensuring the continued supply of electricity. Microgrids are emerging as an effective solutions for supporting power system resilience while providing opportunities to integrate distributed renewable energy generation efficiently into the utility grid during normal operations. The resilience of microgrids against physical, cyber, and communications threats must be ensured by proactively addressing these threats in a holistic way.

This work is focused on developing a quantitative, holistic approach to identify threats to a microgrid, determine vulnerabilities associated with identified threats, and employ mitigation strategies to ensure resilient performance of the microgrid in grid-connected and islanded modes of operation. Physical threats to which microgrids are typically vulnerable include natural hazards, changing climate, and human-induced attacks. Communications threats pertain to an operating condition when the communication between microgrid and utility grid or within a microgrid is unintentionally broken. Threats posed by cyberattacks can range from unintended islanding to triggering out-of-phase reclosing in a microgrid to damage the rotating machines. Due to the highly interdependent nature of Critical Infrastructure (CI) systems, there are threats posed to the microgrid operation from events including interruption in natural gas and/or water supply. The discussion of interdependencies between CI systems and their impact on resilience modeling of microgrid is also presented. The quantitative threat modeling approach is presented to calculate the risk factor.

This paper discusses various mitigation strategies are proposed for pre-, during-, and after-disaster recovery modes. The proposed mitigation strategies include system hardening and operational effectiveness measures. Mitigation strategies are also classified to assist the enhancement of different attributes of the microgrid, including robustness, redundancy, resourcefulness, response, and recovery. Practical consideration in prioritizing the mitigation strategic most impactful for a microgrid in a given situation are also discussed with an example.

The future work will include the development of metrics to quantify the resilience of microgrids to various threats. By quantifying the resilience of different tactics, decision-makers will be able to assess the economic feasibility of technology and operational options and to answer the question, 'is the capital investment required to build resilient microgrid worth the benefits reaped during a probable disruptive event in the future?'. These cost-benefit analyses will aid in the development of policies and practices to guiding microgrid development and deployment.

## VIII. Acknowledgments

This work was authored by the National Renewable Energy Laboratory (NREL), operated by Alliance for Sustainable Energy, LLC, for the U.S. Department of Energy (DOE) under Contract No. DE-AC36-08GO28308. This work was supported by the Laboratory Directed Research and Development (LDRD) Program at NREL. Authors wish to thank Bob Wood and Maurice Martin (NREL) for providing useful suggestions to refine the manuscript. The views expressed in the article do not necessarily represent the views of the DOE or the U.S. Government. The U.S. Government retains and the publisher, by accepting the article for publication, acknowledges that the U.S. Government retains a nonexclusive, paid-up, irrevocable, worldwide license to publish or reproduce the published form of this work, or allow others to do so, for U.S. Government purposes.